\documentclass[twocolumn]{aa}
\usepackage{graphicx}
\usepackage{latexsym,natbib}
\def\cf{cf.~}
\def\eg{e.g.~}
\def\fig{Fig.\,}
\def\sec{Sect.\,}
\def\eq{\!=\!}
\def\ltsim{~\rlap{\lower -0.5ex\hbox{$<$}}{\lower 0.5ex\hbox{$\sim\,$}}}

\def\magsqarcsec{mag$/$\raisebox{-0.4ex}{\hbox{$\Box^{\prime\prime}$}\,}}
\def\rbr{\hbox{$R_{\rm br}$}\,}
\def\hin{\hbox{$h_{\rm in}$}\,}
\def\hout{\hbox{$h_{\rm out}$}\,}
\def\mubr{\hbox{$\mu_{\rm br}$}\,}

\def\kms{km\,s$^{-1}$}

\begin{document}
   \title{Evidence for a Large Stellar Bar in the \\
Low Surface Brightness Galaxy UGC\,7321.  
\thanks{Based on observations obtained at the German-Spanish Astronomical 
Center (DSAZ), Calar Alto, jointly operated by the Max-Planck-Institut f\"ur 
Astronomie Heidelberg and the Spanish National Commission for
Astronomy.}
}

%   \subtitle{}

   \author{M. Pohlen
          \inst{1,2}
          \and
          M. Balcells \inst{1}
%\fnmsep\thanks{Just to show the usage  of the elements in the author field}
          \and 
          R.~L\"utticke \inst{3,2}
          \and 
          R.-J. Dettmar \inst{2}
          }

   \offprints{M. Pohlen}

\institute{
Instituto de Astrof\'{\i}sica de Canarias, 
E-38200 La Laguna, Tenerife, Spain\\
\email{pohlen,balcells@ll.iac.es}
\and
Astronomical Institute, Ruhr-Universit\"at Bochum, 
D-44780 Bochum, Germany\\ 
\email{dettmar@astro.rub.de}
\and
Department of Computer Science, FernUniversit\"at Hagen,
D-58084 Hagen, Germany \\
\email{Rainer.Luetticke@FernUni-Hagen.de}
         }

%             \email{c.ptolemy@hipparch.uheaven.space}
%             \thanks{The university of heaven temporarily does not
%                     accept e-mails}

   \date{Received 5 June, 2003; accepted 8 July, 2003}

   \abstract{ 
Late-type spiral galaxies are thought to be the dynamically simplest 
type of disk galaxies and our understanding of their properties plays
a key role in the galaxy formation and evolution scenarios. 
The {\it low surface brightness} (LSB) galaxy UGC\,7321, a nearby, 
isolated, ``superthin'' edge-on galaxy, is an ideal object to study those 
purely disk dominated bulge-less galaxies.
Although late type spirals are believed to exhibit the simplest 
possible structure, even prior observations showed deviations 
from a pure single component exponential disk in the case of UGC\,7321.
We present for the first time photometric evidence for peanut-shaped
outer isophotes from a deep optical (R-band) image of UGC\,7321.
Observations and dynamical modeling suggest that boxy/peanut-shaped
(b/p) bulges in general form through the buckling instability in bars 
of the parent galaxy disks. 
Together with recent HI observations supporting the presence of a 
stellar bar in UGC\,7321 this could be the earliest known case 
of the buckling process during the evolutionary life of a LSB galaxy, 
whereby material in the disk-bar has started to be pumped up above 
the disk, but a genuine bulge has not yet formed.  
   }
   \maketitle
%
%________________________________________________________________
\section{Introduction}
\label{Introduction}
Nearly all disk galaxies have both a stellar disk and a bulge.
Traditionally, bulges were seen as similar to small ellipticals:
kinematically hot, spheroidal or triaxial, and old. 
Therefore, they probably formed 
during the first collapse of the proto-cloud or
at least in an early stage of their evolution. 
Multicolor studies of highly inclined galaxies 
\cite[eg.][]{balcells1994,peletier1999} support this picture for
early-type disk galaxies: their bulges have ages similar to cluster
ellipticals, though they also tend to be bluer than ellipticals of
similar luminosity.
However, there are galaxies which do not have a bulge component. These 
pure disk galaxies show clearly that the bulge formation is not a 
necessary outcome of the early galaxy formation process. 
Other possible origins of bulges are merger scenarios where either 
a small satellite galaxy is accreted and spirals into the center of the
pre-existing disk \cite[]{aguerri2001} or 
where two equal mass galaxies \cite[]{kauffmann1996} merge and form 
an elliptical structure.
Later, this object again accretes gas, which builds the present 
day surrounding disk.
Another currently favoured scenario is the bulge growth by bar-driven mass
inflow. Starting with a pure disk galaxy, numerical simulations have
shown that the disks are, almost always, unstable against bar formation.
The existing bar is then able to move material into the center. A
buckling vertical instability pumps the stars above the disk, resulting
in a three-dimensional bulge-like object \cite[eg.][]{friedli1993}. The
growing central mass eventually weakens and dissolves the bar 
\cite[]{norman1996,sellwood1999}. Thus we are left with an unbarred 
disk galaxy possessing a spheroidal bulge as observed today. 
In this scenario, pre-existing spheroidal bulges would become contaminated 
by disk stars, which might explain why bulges and the inner parts of disks
have similar colour \cite[]{balcells1994}.
Bars are associated with the appearance of the so called box- or 
peanut-shaped (b/p) ``bulges'' visible in edge-on disk galaxies. 
The b/p part of these bulges are almost certainly the resonant 
off-plane thickening of a bar 
\cite[]{combes1990,raha1991,pfenniger1991,patsis2002a}. 
The connection between bars and b/p bulges has been observationally 
verified by gas kinematics \cite[]{kuijken,bureau1998}, and 
\cite{luett2000b} found a significant correlation between b/p bulges 
and bar signatures seen in cuts along, and parallel to, the major axis 
in NIR images.

\begin{figure*}
\includegraphics[width=10.4cm,angle=270]{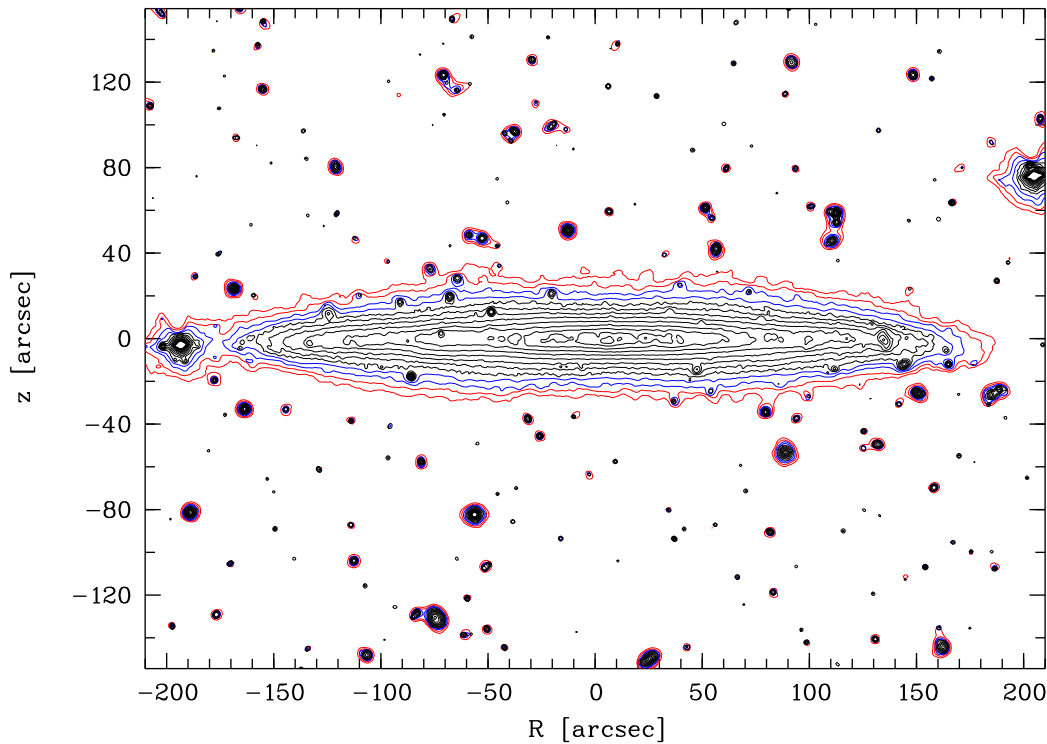}
\caption[]{
Contour map ($\mu_{\rm R}$) of UGC\,7321 from 26.7 to 18.7 equally
spaced by 0.5 mag. \label{cont_big} 
\label{fig1}
}
\end{figure*}

Photometrically observed peanut-shaped contours 
therefore provide evidence for the thickening process caused by the 
buckling instability.
In this paper, we analyze the case of UGC\,7321, a nearly edge-on 
Sd galaxy.  Global properties of this galaxy are given in
Table~\ref{glob_prop}. 
UGC\,7321 does not have a prominent bulge as one can infer from 
its contour map (Fig.~\ref{cont_big}).  However, the outer isophotes in 
deep $R$-band images show peanut-type distortions.  UGC\,7321 provides a 
useful case for studying 
the origin of peanut-type isophotal distortions in 
a galaxy without a prominent bulge. It could provide clues on the very 
early phases of formation of a central dense component. A key question is 
whether the galaxy harbours a bar. UGC\,7321 was studied in detail in 
a series of papers by L.~Matthews \cite[]{matt01, matt02, matt03}.
\cite{matt01} found a red nuclear feature in the $B\!-\!R$ colour map of
UGC\,7321, only a few arcseconds across, and raised the possibility  that it
could be a kinematically distinct disk subsystem analogous to a bulge.  
The existence, definition, and properties of bulges in late-type spiral 
galaxies were recently described by \cite{boeker}. Late-type bulges 
are far from uniform and care needs to be taken regarding what may 
rightfully be called a bulge. Our analysis (\sec{\ref{Results}}) shows 
that, if 
the peanut distortions are indeed the result of a bar, the inferred bar 
length is uncomfortably long as compared to other late-type galaxies.  
However, alternative origins for the peanut distortions, discussed 
in \sec{\ref{Discussion}}, fail to provide convincing explanations.  
\begin{table}
\caption[]{Properties of UGC\,7321. Note: Distance and $M_B$ taken from \cite{matt02}. Other values from RC3 \citep[Third Reference Catalogue of bright galaxies:][]{rc3}. 
  \label{glob_prop}}
\begin{center}
\begin{tabular}{ll}
\hline
\hline
Property                & Value                 \\
\hline
Hubble type            & Sd               \\
T                      & $7.0 \pm 0.8$           \\
$cz_{\rm helio}$ [\kms]& $408 \pm 6$      \\
Distance [Mpc]         & $10 \pm 3$         \\
Scale    [pc/arcsec]   & 48.5          \\
$M_B$    [mag]         & -17.1              \\ 
\hline \\
\end{tabular}
\end{center}
\end{table}
%
%
%
%________________________________________________________________
\section{Data}
\label{Data}
Images of UGC\,7321 were obtained during an observing run in May 2001 at the 
2.2\,m telescope of the Calar Alto observatory (Spain) equipped with 
CAFOS.
The dithered observations of this edge-on galaxy were used as night sky 
flatfields for the original face-on imaging study without spending 
extra dark time and are now discussed here. 
The data reduction is described in detail in \cite{pohlen3} and 
only briefly recalled here. 
The images were taken in a special R-band filter showing a nearly 
rectangular filter characteristic combined with a high peak efficiency. 
Landolt standard fields \cite[]{landolt} were observed to 
achieve a photometric calibration to the standard 
Johnson system \citep[cf.][]{pohlen3}.
During the data reduction process, special emphasis was given to 
performing the crucial flatfielding and sky subtraction.     
The final image used for this study is a combination of six 
stacked 600\,s exposures, which allows one to reliably trace the profiles 
and contours down to $\mu\!\approx 27.2$\,R-\magsqarcsec.
%
%________________________________________________________________
\section{Results}
\label{Results}
\begin{figure}
\includegraphics[width=5.8cm,angle=270]{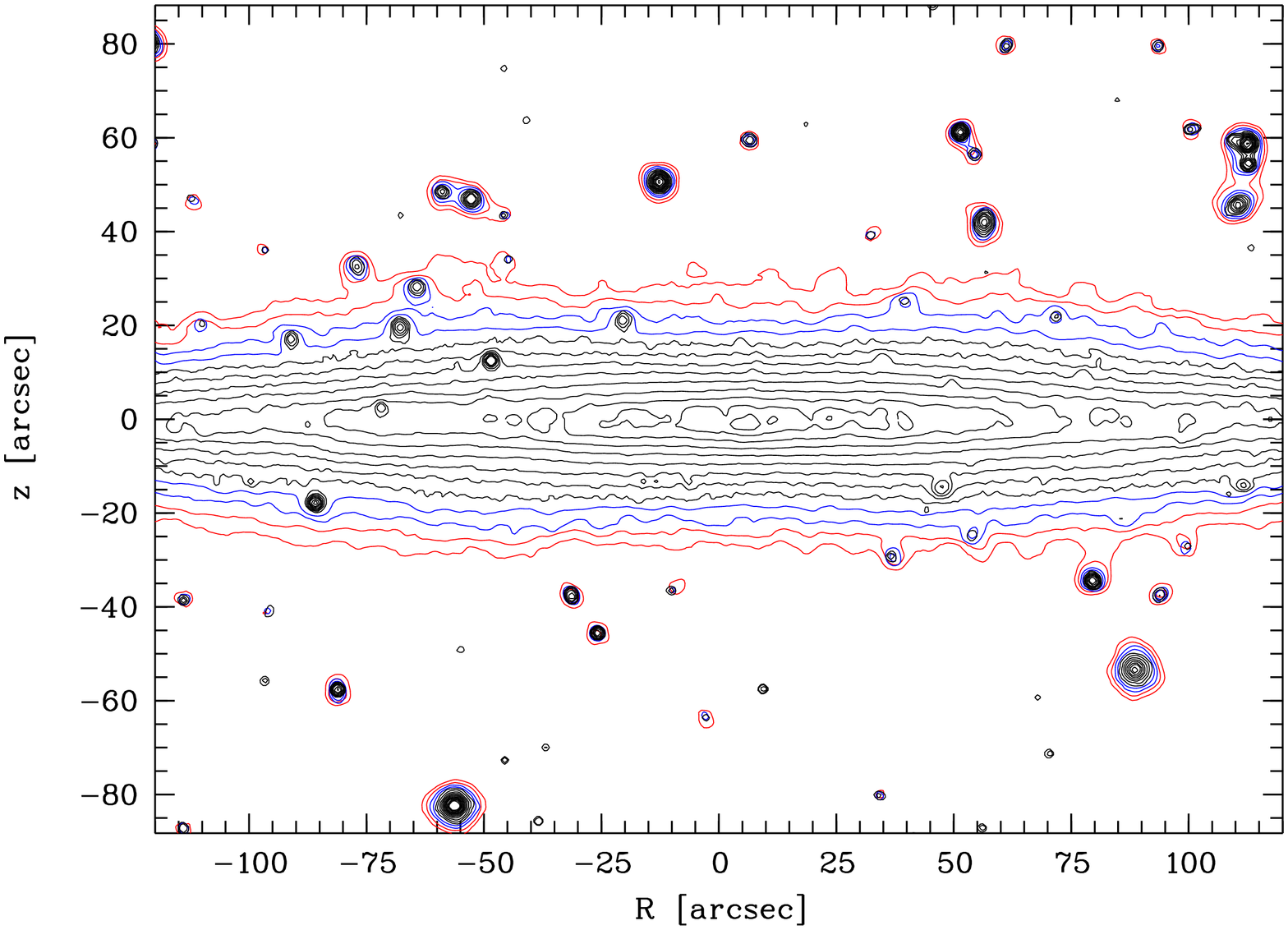}
\includegraphics[width=5.8cm,angle=270]{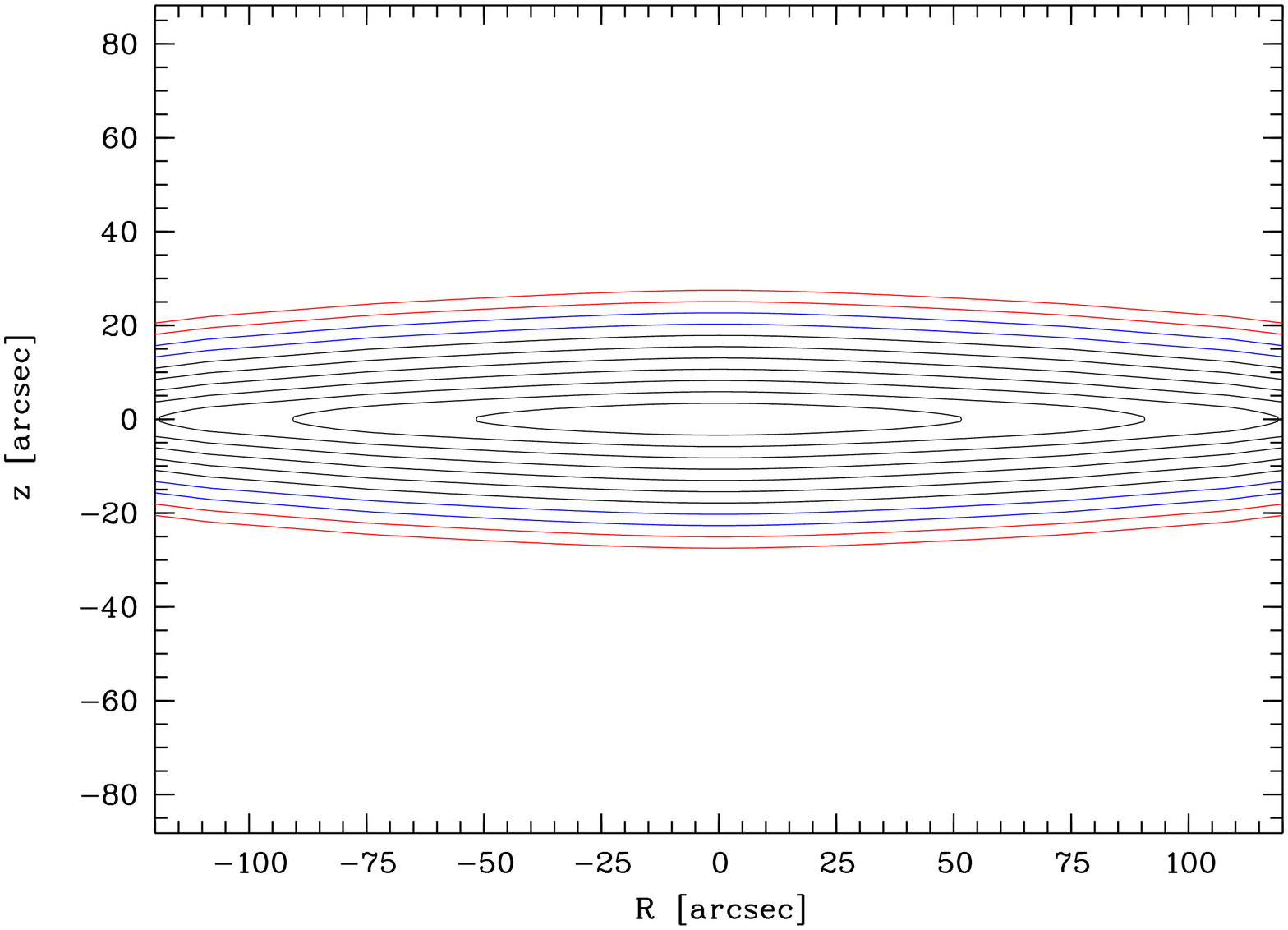}
\caption{
{\sl (Upper panel)} The inner 120\arcsec of UGC~7321 to 
highlight the region of the peanut distortions. {\sl (Lower panel)} 
isophotal contours of the three-dimensional model described in the text.  
Contours from 26.7 to 18.7 are equally spaced by 0.5 mag. 
\label{cont_mod}
}
\end{figure}

\begin{figure}
\includegraphics[width=5.8cm,angle=270]{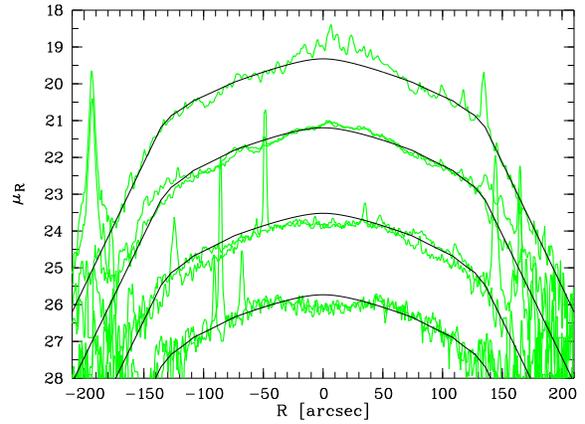}
\caption{ Radial surface brightness profiles of UGC\,7321:  
Major axis (top) and three parallel profiles at $6^{\prime\prime}$, 
$12^{\prime\prime}$, and $18^{\prime\prime}$ above and below the 
major axis. For better visibility the four profiles are 
shifted by $(-2,-1,0,+1)$\,mag, respectively. 
\label{rad}}
\end{figure}

A contour map of the inner 120\arcsec of UGC~7321 ($R$-band) is shown in
\fig\ref{cont_mod}. The six outer isophotes clearly deviate 
from pure ellipticity, with a depression along the minor 
axis above and below the galaxy center.  
According to \cite{luett2000a}, this depression along the minor axis is the 
characteristic feature of a classical, type {\bf 1} peanut-shaped distortion. 
The presence of peanut-shaped distortions is made obvious in cuts
parallel to the major axis.  In \fig\ref{rad} we show such cuts at
0\arcsec, $\pm$6\arcsec, $\pm$12\arcsec and $\pm$18\arcsec.  
The peanut-shape signature (the central "dip" in the profile) is visible in 
the $\pm12^{\prime\prime}$ and obvious in the $\pm18^{\prime\prime}$ 
profiles. The maximum peanut distortion occurs around $\pm50$\arcsec 
from the center. 
The term ``peanut-shaped distortion'' is commonly used to 
describe a property of the galaxy's bulge.  Here, however, 
the galaxy clearly lacks a kinematically
hot, spheroidal $R^{1/4}$ system, and appears to lack a ``pseudobulge'' 
\cite[]{kormendy} or ``exponential bulge'' \cite[]{carollo98}; we show below
that the galaxy has little central light in excess of the best-fit disk
exponential model.  Therefore, the peanut-shaped distortion in UGC~7321 
seems to be a property of the galaxy's disk.
Given the absence of a strong central light concentration, we model the light
distribution of UGC~7321 as a pure disk with the following hypotheses: the
face-on surface brightness profile is exponential with scalelength \hin, out
to a break radius \rbr, and again exponential outside \rbr, with an outer 
scalelength \hout$<$\hin.  The vertical surface brightness profile is also 
taken to be exponential, with scaleheight $z$.  We employ a slightly 
modified version \cite[cf.][]{pohlen5} of the  three-dimensional fitting 
routine described in \cite{pohlen2} to  derive a best-fit model of the 
galaxy's isophotes.  
Radial profiles of the resulting model overplotted on the data are 
shown in \fig\ref{rad}. The best-fit parameters are: inner scalelength 
\hin$\eq79.2$\arcsec, mean break radius \rbr$\eq136.5$\arcsec ($+135.0$\arcsec
and $-138.0$\arcsec), outer scalelength \hout$\eq15.5$\arcsec, and vertical 
scaleheight $z\eq19.7$\arcsec. Assuming a distance of 10.0\,Mpc, this yields 
\rbr$\eq6.6$\,kpc, \hin$\eq3.8$\,kpc, and $z\eq1.0$\,kpc. 
The three-dimensional modeling allows for a rather precise determination 
of the galaxy's inclination in nearly edge-on views. We derive an 
inclination of $i\eq88.5\degr$, a central surface brightness of 
$\mu_0\eq21.3$\,R-\magsqarcsec and $\mubr\eq23.5$\,R-\magsqarcsec 
at the break radius. 
The corresponding face-on values are: $\mu_0\eq23.8$\,R-\magsqarcsec and 
$\mubr\eq25.7$\,R-\magsqarcsec.
A contour plot of the model light distribution is shown in
\fig\ref{cont_mod}. Comparison of the two panels in \fig\ref{cont_mod}
shows that the standard horizontal- and vertical-exponential disk model lacks
the outer peanut distortions seen in UGC\,7321.  Figure\,\ref{rad} 
shows model cuts parallel to the major axis overplotted on the 
corresponding cuts for UGC\,7321.  Here again, the peanut distortions 
in the data  at $R\ltsim 50$\arcsec\ deviate from the model in the lowest, 
outermost two cuts ($z\eq\pm12$\arcsec, $z\eq\pm18$\arcsec).
Two further aspects are worth emphasizing.  First, the good match in the 
outer parts between data and model provides evidence that UGC\,7321, like 
other LSB galaxies (\cite{pohlen4}), has the truncated, two-slope disk 
structure found in high-surface brightness spirals \cite[cf.][]{pohlen3}.
Second, the major-axis cut (top profile in \fig\ref{rad}) shows a central 
enhancement over the best-fit exponential model at $\pm40$\arcsec. This 
structure indicates that an additional component exists besides the 
double-exponential light distribution in UGC\,7321.  
This component is affected by the in-plane dust extinction on the order 
of 0.4\,mag \cite[]{matt03}. Using a simple-minded definition of bulges 
as any excess 
light over the inward extrapolation of the exponential disk profile 
\cite[e.g.][]{carollo99}, we could call the inner excess ``the bulge" of 
UGC\,7321.  Interestingly, the radial position of the peanut 
distortions ($\approx\!\pm50$\arcsec) is comparable to the radial
extent of the inner brightening, suggesting a parallel with the 
isophotal structure of classical box-peanut bulges  \cite[cf.][]{luett2000b}.
Alternatively, the inner brightening of the major-axis profile might trace 
a bar in UGC\,7321. Its profile and radial extent are hard to infer from the
major-axis profile given the extinction by dust. The properties of such a 
bar might be inferred from the peanut distortions if we assume that the
peanut feature is produced by the resonant off-plane thickening of 
a bar (\eg Combes et al.\ 1990). 
Following \cite{luett2000b}, we derive the length of the peanut structure 
({\bf BPL}$=\!83$\arcsec) as the radial distance between the maxima of the 
peanut distortion (in detail: $-39$\arcsec\ and $+44$\arcsec) measured in 
radial cuts. 
For a sample of normal HSB galaxies, \cite{luett2000b} find a mean value
of $2.7\pm0.3$ for the ratio of projected  bar length ({\bf BAL}) to {\bf
BPL}.  Assuming a similar scaling, the projected length of the bar in
UGC\,7321 would be $\approx224\pm 25$\arcsec\ in diameter and 
$112 \pm 13$\arcsec\ in radius.
The two extreme values for the same sample 
({\bf BAL}/{\bf BPL}$=\!2.2$ or $3.2$)
would allow projected bar radii in the range of $91$\arcsec\ to $133$\arcsec.
The estimated bar length is quite large. It is comparable to the total galaxy
diameter, and would indicate that most of the body of UGC\,7321 comprises a 
bar seen side-on with a ratio of bar length to galaxy size of {\bf
BAL}/$D_{25}\eq 224/302\eq 0.7$ \cite[$D_{25}\!\equiv\!$ observed 
25\magsqarcsec B-band diameter uncorrected for inclination, from][]{matt01}.  
In comparison, for a sample of normal disk galaxies observed
in the NIR, \cite{luett2000b} derive a mean value of {\bf BAL}/$D_{25}\eq
0.4\!\pm\!0.2$. $D_{25}$ is probably not a good indicator for comparing LSBs 
with normal disk galaxies, but any inclination correction for $D_{25}$ will
increase the ratio {\bf BAL}/$D_{25}$ even more. 
The numbers given above depend somewhat on the precise definition used for bar
length and peanut length. \cite{luett2000b} define the beginning of the bar 
in radial profiles as the location where the bar extends above the 
extra\-polated outer disk. 
If bars are defined to start at their shoulders or bumps, {\bf BAL} will 
be a factor of 1.3 smaller. 
In addition, using the maximum of the peanut rather than the full 
structure, {\bf BPL} will be also a factor of 1.2 
larger. This results in {\bf BAL}/{\bf BPL}$=\!2.1$ or even 
{\bf BAL}/{\bf BPL}$=\!1.7$, which is 
closer to the value obtained in N-body simulations \cite[]{pfenniger1984}.
For any of the given definitions, the case remains that the inferred bar 
length for UGC\,7321 is quite large.  
Comparing {\bf BAL} with the radial scalelength ($h$) instead yields a 
ratio of {\bf BAL}/$h\eq224/79\eq2.8$. Even assuming the lowest  
measured {\bf BAL}/{\bf BPL} ratio of $2.2$ from \cite{luett2000b}, 
{\bf BAL}/$h$ will still be $182/79\eq2.3$.
This is comparable to a bar of an early type galaxy. 
\cite{erwin} derives a mean value of {\bf BAL}/$h\eq2.4\!\pm\!1.0$ for a 
sample of face-on S0 to Sab galaxies and obtained a ratio of 
{\bf BAL}/$h\eq1.2\!\pm\!0.5$ for Sd galaxies taken from a sample 
of \cite{martin}. 
Another example is the edge-on Sb galaxy NGC\,2424. It exhibits a ratio 
of {\bf BAL}/$h\eq2.6$, taking {\bf BAL} $\eq76$ from \cite{luett2000b}
and $h\eq29$ from \cite{pohlen5}\footnote{
The scalelength $h$ is determined only 
one-di\-men\-sionally and is therefore just a lower limit. In the worst 
case it is off by a factor of $+20$\% compared to the 
true 3D scalelength 
\cite[cf.][]{pohlen5}, which allows a {\bf BAL}/$h$ as high as 3.3 
for NGC\,2424.}.
Note that the
apparently unknown orientation of a bar in an edge-on galaxy is 
restricted by the actual peanut appearance of the contours. \cite{luett2000b} 
have shown that the bar-produced b/p structure appears peanut-shaped only for 
a small range of aspect angles between $77\degr-90\degr$ for the associated 
bar, assuming a galaxy inclination of $90\degr$.
UGC\,7321 is classified as a non box/peanut-shaped (b/p) bulge 
(type {\bf 4}) in the catalogue of \cite{luett2000a} and is actually the
first Sd galaxy with a peanut bulge. 
However, they classify ``bulges'' to be either elliptical ({\bf 4}), 
boxy ({\bf 3+2}), or peanut-shaped ({\bf 1}), and therefore concentrate
only  
on the inner part of UGC\,7321 (which is indeed type {\bf 4})
visible in DSS (Digitized Sky Survey) images.  
The peanut structure here is much larger than for a typical
galaxy with a b/p-bulge. 
In addition, the characteristic angle $\theta$ between the major axis and 
the radial ray passing through the maximum of the peanut distortion is 
much smaller than normal \cite[cf.][]{luettdiss,shaw}. For UGC\,7321, $\theta$
is $\approx18\degr$ while the mean value for all galaxies with peanut-shaped 
bulges is $39\degr\!\pm\!10\degr$ with a minimum 
of $31\degr$ \cite[]{luettdiss}.
The conclusion would be that we are seeing an edge-on bar 
without the additional (superimposed) spheroidal bulge component as 
is typical for
earlier (S0-Sc) galaxy types. Therefore, this large, slightly peanut-shaped, 
outer structure is {\sl the bar} and one has to consider that 
the term b/p-bulge is a combination of a b/p structure, which we have here, 
and a spheroidal bulge component. 
%   

%________________________________________________________________
\section{Discussion} 
\label{Discussion}
Inferring the presence of a bar from the weak isophotal distortion 
observed in UGC\,7321 might be controversial, especially given 
the implied large bar size compared to the galaxy disk. Here, we examine 
what other evidence exists for or against the presence of the bar, and 
what other configurations may create the observed peanut distortion.
Strong support for our photometric bar detection comes from the 
recent analysis of HI-data for UGC\,7321 by \cite{uson}. They find 
that the major axis position-velocity (P-V) profile shows a clearly 
visible ``figure-of-eight'' structure (\cf their \fig 10). 
This is typical for a barred galaxy viewed edge-on \cite[]{kuijken}.
\cite{uson} favour the explanation proposed by \cite{brinks} that 
this feature may also result from the warping and flaring of its HI disk.
However, there is only a very mild --- compared to other later type galaxies 
\cite[]{garcia} --- HI warp on the west side\footnote{here: positive side}.
\cite{uson} assign a decrease in the HI intensity at $\approx-25\arcsec$ 
to the possible presence of a small bar extending from about $-30\arcsec$
to $+40\arcsec$ which nicely coincides\footnote{Note: We use a different 
orientation of the image.} with our ``bulge'' feature in the major
axis light profile.
Nevertheless, if this is a bar it would be too small to account for 
the peanut-shaped outer structure. 
We do not find indications of a flaring in the 
{\sl stellar} disk. 
By introducing a flaring in our model we confirm that it could not be 
responsible for the observed peanut distortion in the inner part. 
The effect of a radially increasing scaleheight is most 
prominent towards the edges of the disk.  
\cite{binney} and \cite{whitmore} have shown that the accretion of satellite 
galaxies is a possible formation process for b/p bulges. 
For normal b/p-bulges these scenarios are excluded due to
the overwhelming  evidence for bar-driven evolution 
\cite[eg.][]{bureau2002,luett2000b}, where the role of interaction is 
only to trigger the bar itself. 
However, \cite{luett2003} found a special class of thick boxy bulges (TBB) 
which suggest a merger origin. These galaxies are often disturbed, 
frequently showing prominent irregularities and asymmetries, and 
possess significantly more projected satellites compared to a control sample.
All these characteristics are in contrast to UGC\,7321's smooth and 
undisturbed disk and its unusually isolated environment. There 
are no signs for a recent merger event, such as 
an increase in star-formation, asymmetric
dust distribution, arcs, shells, filaments, or a strong 
stellar warp. 
\cite{uson} estimate that a possible encounter $\approx\!10^9$ years ago 
of UGC\,7321 with its closest neighbour could be responsible for the
observed mild HI warp. 
Taken together, nearly all of the observational evidence points
against a merger scenario. The peanut-shaped structures found in
simulations \cite[eg.][]{hernquist} resemble typical early-type
spirals and S0s such as NGC\,128, IC\,4767, or HCG\,87A and are often located
in dense groups; some of them are even ``hybrid'' scenarios where
the bar is triggered by a merger \cite[eg.][]{mihos}.
\cite{patsis2002a}, analysing N-body models, describe a new vertical 
resonance producing an edge-on `boxy' or `peanut' structure for 
non-barred galaxies. However, the examples from the N-body 
simulation have only an oval distortion as their main morphological 
feature and could be characterized morphologically as S0 galaxies.
In addition, their model allows boxes, but not strong peanuts
or X-shaped structures. 
Therefore, the presence of a large bar 
extending over the main body of the 
galaxy provides the best explanation for the peanut-shaped contours 
in UGC\,7321 and the measured ``figure-of-eight'' pattern in the 
recent HI measurements.
A bar, especially one as big as we photometrically infer, in a low
surface brightness galaxy, a system expected to be dark-matter
dominated, could solve the problem that many rotation
curves of LSB galaxies cannot \cite[eg.][]{deblok} be fitted by the
cuspy halo profiles predicted by CDM cosmogonies. Specifically,
\cite{swaters} discuss non-circular, bar-induced kinematics as the
source of deviations between rotation curve-based mass models 
and the NFW-type CDM halos.
\cite{weinberg} propose that a dark matter cusp might be removed due to
dynamical interaction of the halo with the bar.
However, several works argue against such a mechanism and in 
favor of non-circular kinematics to explain the deviations. For example, 
\cite{mayer} used high resolution N-Body/SPH simulations to
address the question of bar formation and evolution in LSB
galaxies. They conclude that LSB disks may go bar-unstable
in gas-rich disks embedded in low-concentration halos. However, the
resulting bars are too short and probably of too low angular momentum
content to affect the inner halo density profile.  They also conclude
that a bulge-like component must be present in any LSB galaxy 
that has got bar unstable.
In the case of UGC\,7321 the inner component could
be tracing this process,
so that we are observing just the beginning of
the morphological evolution from its first bar into a bulge-like
central structure, according to the secular evolution scenario.
%
%________________________________________________________________
\begin{acknowledgements}
Part of this work was supported by the German
\emph{Deut\-sche For\-schungs\-ge\-mein\-schaft, DFG}. 
We would like to thank L.~Matthews for stimulating 
discussions about UGC\,7321 and D.~Pfenniger and P.~Patsis for their 
views about a possible bar scenario.  
\end{acknowledgements}
%________________________________________________________________

\end{document}